\documentclass[aps, prfluids, reprint, superscriptaddress, longbibliography, floatfix, amsmath, amssymb]{revtex4-2}

\usepackage{graphicx,dcolumn,bm}
\usepackage[nameinlink,capitalise]{cleveref}
\usepackage[dvipsnames]{xcolor}

\crefname{equation}{Eq.}{Eqs.}
\crefname{figure}{Fig.}{Figs.}
\crefname{table}{Table}{Tables}
\crefname{section}{Sec.}{Secs.}
\crefname{subsection}{Sec.}{Secs.}
\crefname{subsubsection}{Sec.}{Secs.}
\crefname{appendix}{Appendix}{Appendices}

\newcommand{\singlefigwidth}{\columnwidth}

\begin{document}

\title{Contact-network organization and motion statistics in shear-thickening suspensions}

\author{Michel Orsi}
\email{Contact author: michel.orsi@polito.it}
\affiliation{Benjamin Levich Institute, CUNY City College of New York, New York, NY 10031}
\affiliation{Department of Applied Science and Technology, Politecnico di Torino, 10129, Torino, Italy}

\author{Rahul Pandare}
\affiliation{Benjamin Levich Institute, CUNY City College of New York, New York, NY 10031}

\author{Brolin Adu-Poku}
\affiliation{Benjamin Levich Institute, CUNY City College of New York, New York, NY 10031}

\author{Bulbul Chakraborty}
\affiliation{Martin Fisher School of Physics, Brandeis University, Waltham, MA 02454}

\author{Jeffrey F. Morris}
\affiliation{Benjamin Levich Institute, CUNY City College of New York, New York, NY 10031}
\affiliation{Department of Chemical Engineering, CUNY City College of New York, New York, NY 10031}

\date{\today}

\begin{abstract}
We use lubricated-flow discrete-element-method (LF--DEM) simulations to examine how contact-network organization shapes particle motion in dense shear-thickening suspensions. The primary system studied is a two-dimensional bidisperse monolayer where rigid clusters are identified by the $(3,3)$ pebble game; three-dimensional simulations are shown to have qualitatively similar rotational velocity statistics. Across the stress--solid-fraction state diagram, frictional contact number, $k\ge 3$ percolation, and rigid-cluster fluctuations all strengthen in the same region where translational velocity correlations grow, consistent with coherent translation within rigid clusters and a state-dependent contrast in relative motion between particles inside and outside them. Rotational motion provides a complementary view: non-affine angular-velocity distributions broaden, near-contact non-affine rotational fluctuations become increasingly anti-correlated, and particles inside and outside rigid clusters carry distinct statistics. Connectivity, rigidity, and velocity correlations are related but distinct signatures of the constrained collective motion that accompanies shear thickening and the approach to shear jamming.
\end{abstract}

\maketitle

\section{Introduction}\label{sec:intro}

Dense non-Brownian suspensions appear in many natural and industrial flows. Their rheology is simple only at first sight: even when the suspending liquid is Newtonian and inertia is negligible, particles can generate strong normal stresses resulting in particle migration, and rate-dependent effects, of which the most striking example is shear thickening \citep{denn2014,morris2020,morris2023,morris2025}. In discontinuous shear thickening (DST), the viscosity rises rapidly once the applied stress is large enough, and at sufficiently high solid fraction, $\phi$, the flowing state can approach shear jamming \citep{brown2012,brown2014,seto2013,mari2014,mari2015,wyart2014,morris2018}.

A useful way to understand this transition is to view stress as a switch that changes the allowed interactions. At low stress, lubrication and repulsive forces keep particles from making sustained frictional contact. At larger stress, the repulsive barrier is overcome and contacts become possible \citep{seto2013,mari2014,mari2015,mari2015a,wyart2014,guy2015,guy2018,singh2018,singh2020,singh2022}. Experiments and simulations have refined this picture by showing how surface roughness, tribology, adhesion, hydrodynamics, contact forces, and particle-size distribution affect the onset and strength of thickening \citep{lin2015,blanc2018,gallier2014a,hsu2018,hsu2021,thomas2018,malbranche2023}. Scaling approaches have also shown that much of the rheology can be organized by the crossover from lubricated or frictionless interactions toward a frictionally dominated state  \citep{malbranche2022,ramaswamy2023,ramaswamy2025}.

A mean contact number captures how many frictional constraints are present on average, but not how they are arranged. Contact and force networks carry spatial structure, can percolate, and are heterogeneous in time and space \citep{edens2021,sedes2022,goyal2024,sharma2026,moghimi2024,moghimi2025}. While observation of such structures has primarily been based on numerical simulation, experimental work has found localized or transient jammed regions and complex flow heterogeneity in thickening suspensions \citep{rathee2020,barik2024,moghimi2024,moghimi2025}. These observations suggest that the relevant microstructure is not a local one such as the pair distribution or contact count, but is a network of constraints that organize motion over mesoscopic distances.

Rigidity provides one way to make this idea more precise. A cluster is called minimally rigid when its frictional contacts reduce the internal degrees of freedom to those of a single rigid body; because the particles used in our simulations are slightly deformable, the identified clusters exhibit collective behavior without being perfectly rigid. Pebble-game methods identify such clusters efficiently and have been useful in describing frictional jamming and related behavior in other disordered systems \citep{jacobs1997,henkes2016,ellenbroek2015,lester2018,liu2019,koeze2018,dashti2023}. Recent applications to shear-thickening suspensions have shown that rigid clusters can appear below shear jamming, that their fluctuations define a critical line in the stress--solid-fraction plane, and that the transition persists in polydisperse systems \citep{santra2025,pandare2026,vandernaald2024,singh2026}. Thus, rigidity is an outcome of the contact network.

The question we address here is how these constraint-based structures are reflected in particle motion. Translational fluctuations, shear-induced diffusion, and velocity correlations are related to the microstructure in sheared suspensions \citep{leighton1987a,brady1997,drazer2004,pham2015,souzy2015,singh2023,athani2024}. Rotational motion provides a complementary probe because tangential constraints and rolling or sliding at contacts couple particle rotations to the surrounding contact network; rotational and torque fluctuations have also been discussed for granular systems \citep{singh2020,singh2022,rahbari2021,zhang2024}. We therefore examine translational and rotational statistics, and show that their changes occur in the same part of the state diagram where we find strong growth in  frictional contact number, percolation of particles with $k\ge 3$ frictionally-contacting neighbors, and rigid-cluster size  fluctuations.

We use stress-controlled lubricated flow-discrete element method (LF--DEM) simulations of dense shear-thickening suspensions. We focus our analysis on two-dimensional bidisperse monolayers, where the frictional contact network is analyzed with a pebble-game rigidity construction; a limited set of three-dimensional simulations demonstrates that the rotational statistics of 2D and 3D suspensions are essentially similar, providing confidence that the primary conclusions related to the kinematic effects of frictional constraints are not limited to 2D. Structural and kinematic observables are mapped onto the well-known flow-state diagram \cite{wyart2014,morris2020} for shear-thickening suspensions to identify where changes in contact-network organization and particle motion are most strongly correlated.

\section{Numerical method}\label{sec:method}

We follow the LF--DEM formulation used in prior work (\citet{seto2013,mari2014,santra2025,pandare2026}). Most results are for a two-dimensional (2D) monolayer of neutrally buoyant, non-Brownian bidisperse spheres with size ratio $a_l/a_s=1.4$ in stress-controlled simple shear with Lees-Edwards periodic boundary conditions \citep{lees1972computer}. This level of bidispersity maintains amorphous structure, avoiding the formation of ordering into strings in the shear flow direction. In the 2D case, $\phi$ denotes the particle area fraction. We also use a limited set of three-dimensional simulations for comparison, with $\phi$ then denoting particle volume fraction. The planar geometry is the primary focus because the rigidity analysis below is well-defined on two-dimensional contact networks \citep{henkes2016,santra2025,pandare2026,singh2026}.

The particle motion is overdamped and satisfies force and torque balance,
\begin{align}
    \mathbf{F}_{\mathrm{H}}(\mathbf{x},\mathbf{U}_{\mathrm{P}}) + \mathbf{F}_{\mathrm{R}}(\mathbf{x}) + \mathbf{F}_{\mathrm{C}}(\mathbf{x},\mathbf{U}_{\mathrm{P}}) &= \mathbf{0}\ , \\
    \mathbf{T}_{\mathrm{H}}(\mathbf{x},\mathbf{U}_{\mathrm{P}}) + \mathbf{T}_{\mathrm{C}}(\mathbf{x},\mathbf{U}_{\mathrm{P}}) &= \mathbf{0}\ .
\end{align}
Hydrodynamic forces are written as
\begin{align}
    \mathbf{F}_{\mathrm{H}} &= -\mathcal{R}_{\mathrm{FU}} \cdot (\mathbf{U}_{\mathrm{P}}-\mathbf{U}_{\infty}) + \mathcal{R}_{\mathrm{FE}} : \mathbf{E}_{\infty}\ , \\
    \mathcal{R}_{\mathrm{FU}} &= \mathcal{R}_{\mathrm{Stokes}} + \mathcal{R}_{\mathrm{Lubrication}}\ ,
\end{align}
where the pairwise lubrication resistances are those of \citet{jeffrey1984} and \citet{jeffrey1992}. Direct contacts follow a Coulomb friction law \citep{cundall1979discrete} with friction coefficient $\mu=100$. This artificially large value is used so that contacts are strongly frictional, with infrequent sliding contacts; the phenomenology is similar to the case of $\mu =O(1)$ with larger viscosity rise at the shear thickening and jamming fraction $\phi_{\rm J}$ shifting to smaller values with increase of $\mu$, as shown by \citet{mari2014}. No independent rolling-resistance law is included. A short-range electrostatic repulsion is modeled to prevent contact interactions until sufficient stress is imposed,
\begin{equation}
    \mathbf{F}_{\mathrm{R}} =
    \begin{cases}
        \mathbf{F}_0 \, D^* \exp(-d/\lambda)\ , & d \ge 0\ , \\[4pt]
        \mathbf{F}_0 \, D^*\ , & d < 0\ ,
    \end{cases}
\end{equation}
with $\mathbf{F}_0$ directed along the line of centers, $D^*=2a_i a_j/[a_s(a_i+a_j)]$, $d=r-a_i-a_j$, and $\lambda=0.05a_s$ \citep{israelachvili2011intermolecular,durand2000surprisingly,zeng2011effect,santra2025,pandare2026}. The repulsive scale defines $\dot{\gamma}_0=F_0/(6\pi\eta_0 a_s^2)$ and $\sigma_0=F_0/(6\pi a_s^2)$, where $F_0=\lvert \mathbf{F}_0\rvert$. Contact occurs when the shearing force driving a pair together exceeds $F_0$, which becomes probable when $\sigma/\sigma_0>1$.

Under stress control, the instantaneous shear rate is obtained from \cite{mari2015}
\begin{equation}
    \dot{\gamma} (t) = \dfrac{\sigma-\sigma_r-\sigma_c}{\eta_0(1+2.5\phi)+\eta_h}\ ,
\end{equation}
with $\sigma_r$ and $\sigma_c$ the repulsive and contact stress contributions, and
\begin{equation}
    \eta_h = V^{-1} \left\{ \left( \mathcal{R}_{\mathrm{SE}} - \mathcal{R}_{\mathrm{SU}} \cdot \mathcal{R}_{\mathrm{FU}}^{-1} \cdot \mathcal{R}_{\mathrm{FE}} \right) : \mathbf{E}_{\infty} \right\}_{12}\ ,
\end{equation}
where $V$ is the simulation volume, or area in 2D, containing $N_p$ particles.
The simulations are inertialess and non-Brownian, with Reynolds number $Re = 0$ and P\'eclet number $Pe \to \infty$. From statistically steady simulations, we extract relative viscosity $\eta^s$, contact-network connectivity $Z_{\mathrm{net}}$, rigid-cluster statistics, and translational and rotational velocity correlations and statistics. All 2D results use $N_p=2000$ particles, with averages taken over steady-state configurations sampled after the accumulated shear strain exceeds $\gamma=2$.

\subsection{Rigid clusters}\label{sec:pebblegame}

Rigid clusters are identified from the frictional contact network using the $(3,3)$ pebble-game (PG) algorithm \citep{jacobs1997,henkes2016,lester2018}, following its application to suspensions by \citet{vandernaald2024,santra2025,pandare2026,singh2026}. In 2D, each unconstrained particle has three (two  translational, one rotational) degrees of freedom, and thus each node in the network, representing a particle center, starts with three pebbles. A sliding contact removes one degree of freedom and a non-sliding frictional contact removes two; a connected subgraph with only three residual degrees of freedom is then classified as minimally rigid.  Because the PG applied here counts frictional constraints, the resulting objects should be interpreted as minimally rigid clusters rather than perfectly rigid bodies \citep{liu2019,vandernaald2024,pandare2026}; note also that the PG rigidity is a definition based upon a degrees of freedom analysis and does not imply true rigidity because our particles have some softness, with time-dependent deformation at contacts providing a mechanism for motion and hence continued flow. The latter point has been  discussed more in other work \cite{pandare2026}. Because this PG construction applies only to 2D contact networks, the rigidity analysis is restricted to the monolayer simulations. The 3D cases are used only for observables that do not require a pebble-game decomposition. 

\begin{figure*}[!t]
    \centering
    \includegraphics[width=\textwidth]{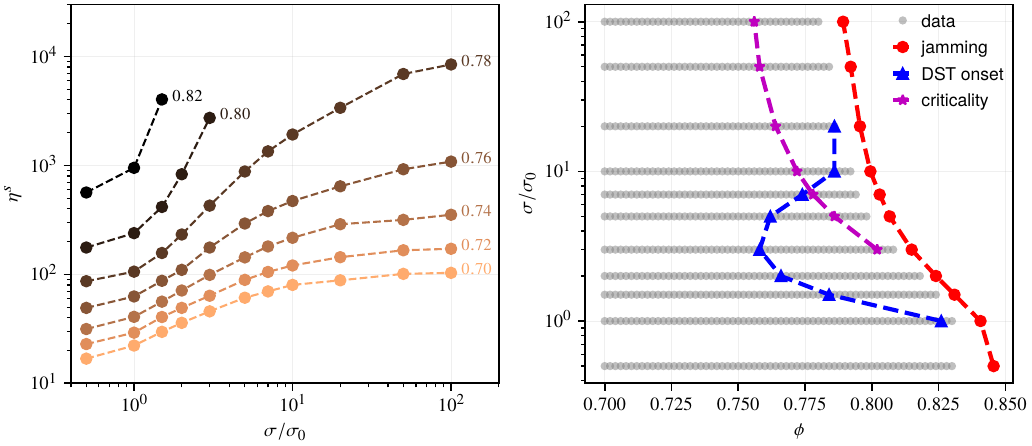}
    \caption{Flow curves and state diagram for the 2D monolayer simulations. Left: stress-dependent relative viscosity $\eta^s(\sigma/\sigma_0,\phi)$ for a representative subset of solid fractions. Right: flow diagram in the $(\sigma/\sigma_0,\phi)$ plane; gray circles denote simulated state points. Red circles mark the shear-jamming boundary $\phi_J(\sigma)$, obtained from Maron--Pierce fits of $\eta^s(\phi)$ at fixed stress; blue triangles mark the DST onset, identified by the onset of $d\langle\dot{\gamma}\rangle/d\sigma<0$ in the stress-controlled flow curves; and magenta stars mark the rigidity critical line $\phi_c(\sigma)$, located from the maximum of $\chi_{\mathrm{rig}}$ at fixed stress.}
    \label{fig:flow_diagram}
\end{figure*}

We use an order parameter based on the rigid structures found. Following \citet{santra2025}, we define $n_i=1$ if particle $i$ belongs to a rigid cluster whose contacts are all internal to that cluster, and $n_i=0$ otherwise. The instantaneous rigid fraction is
\begin{equation}
    m_{\mathrm{rig}} = \frac{1}{N_p}\sum_{i=1}^{N_p} n_i\ .
    \label{eq:mrig}
\end{equation}
We denote its mean by $\langle m_{\mathrm{rig}}\rangle$; in our previous work \citep{santra2025,pandare2026} this mean was denoted $f_{\mathrm{rig}}$. We also determine the susceptibility
\begin{equation}
    \chi_{\mathrm{rig}} = N_p\left(\langle m_{\mathrm{rig}}^2\rangle - \langle m_{\mathrm{rig}}\rangle^2\right) ,
\end{equation}
together with the mean frictional contact number $Z_{\mathrm{net}}$ of the contact network \citep{santra2025,pandare2026}. These observables separate connectivity from rigidity. The contact number measures how many frictional constraints are present, while $\langle m_{\mathrm{rig}}\rangle$ and $\chi_{\mathrm{rig}}$ quantify how much of that network belongs to rigid domains and how strongly the rigid fraction fluctuates. In these 2D suspensions, earlier work found that rigid clusters emerge at a critical $\phi_c (\sigma)$ below shear jamming and become increasingly system-spanning as jamming is approached \citep{vandernaald2024,santra2025,pandare2026,singh2026}. The function $\phi_c(\sigma)$ defines a line of critical points, where $m_{\rm rig}$ fluctuates most strongly (Supplemental Material, Fig.~S7) and thus $\chi_{\rm rig}$ is maximized; this line is labeled as `criticality' in the right panel of \cref{fig:flow_diagram}. \citet{pandare2026} found that, for sufficiently stiff particles, fluctuations to large instantaneous values of $m_{\mathrm{rig}}$ can result in jamming, suggesting that shear jamming even for realistic stiffness (as in experiments with silica particles) is a second-order transition cut off by large-scale rigidity fluctuations. For the softer particles used here, flow persists over a small window of $\phi>\phi_c$ before jamming.

\section{Flow curves and state diagram}\label{sec:flow}

\cref{fig:flow_diagram} provides the rheological map for materials studied. Note that in the left panel of \cref{fig:flow_diagram},
only a fraction of the volume fractions studied are shown for clarity, with the state points for which data was obtained (and used for development of further results such as contour plots) shown by the gray points in the right panel of \cref{fig:flow_diagram}. Throughout, stress is normalized by the repulsive stress scale $\sigma_0$. At fixed $\phi$, the flow curves show relatively weak thickening at lower solid fraction and a much sharper increase in viscosity at larger $\phi$. The largest $\phi$ values studied approach the shear-jamming boundary. The behavior displayed in \cref{fig:flow_diagram} is consistent with the lubricated-to-frictional scenario: Stokes drag and short-range lubrication dominate at lower stress $\sigma/\sigma_0<1$, while larger stress activates a growing population of frictional constraints that give rise to shear thickening and jamming \citep{brown2012,brown2014,denn2014,seto2013,mari2014,mari2015,wyart2014,guy2015,guy2018,blanc2018,thomas2018,morris2018,morris2020,morris2023,morris2025,singh2018,singh2020,singh2022}.

The right panel of \cref{fig:flow_diagram} places all simulated states in the $(\sigma/\sigma_0\,,\,\phi)$ plane and provides the common frame for the structural and kinematic observables discussed below. The DST onset at each stress is the smallest $\phi$ at which $\langle\dot\gamma\rangle$ first decreases with increasing $\sigma$. The shear-jamming line gives $\phi_J(\sigma)$, extracted by fitting $\eta^s(\phi)$ at fixed $\sigma$ to the Maron--Pierce form $\eta^s \propto (1-\phi/\phi_{\rm J})^{-2}$. The rigidity-onset line,  $\phi_c(\sigma)$, lies at $\phi$ below jamming and is identified, following \citet{santra2025} and \citet{pandare2026}, from the peak of the rigid-cluster susceptibility $\chi_\mathrm{rig}$ at fixed stress: this peak marks a continuous transition in the order parameter $m_\mathrm{rig}$ from a state of finite, non-spanning rigid clusters to one with a system-spanning rigid network. As noted, this  behavior at $\phi_c$ is thus a form of criticality, and is labeled as such in \cref{fig:flow_diagram}.  The following sections use this diagram, in particular by overlaying it with data based on sampling from simulations, to identify the relationships between contact-network changes, kinematic changes, and the rheological state of the material. Three-dimensional (3D) simulations display the same qualitative rheological behavior and state-diagram organization (Supplemental Material, Fig.~S1), as has been shown in prior work \cite{morris2020}.

\section{Contact number}\label{sec:znet}

\begin{figure}[!t]
    \centering
    \includegraphics[width=\singlefigwidth]{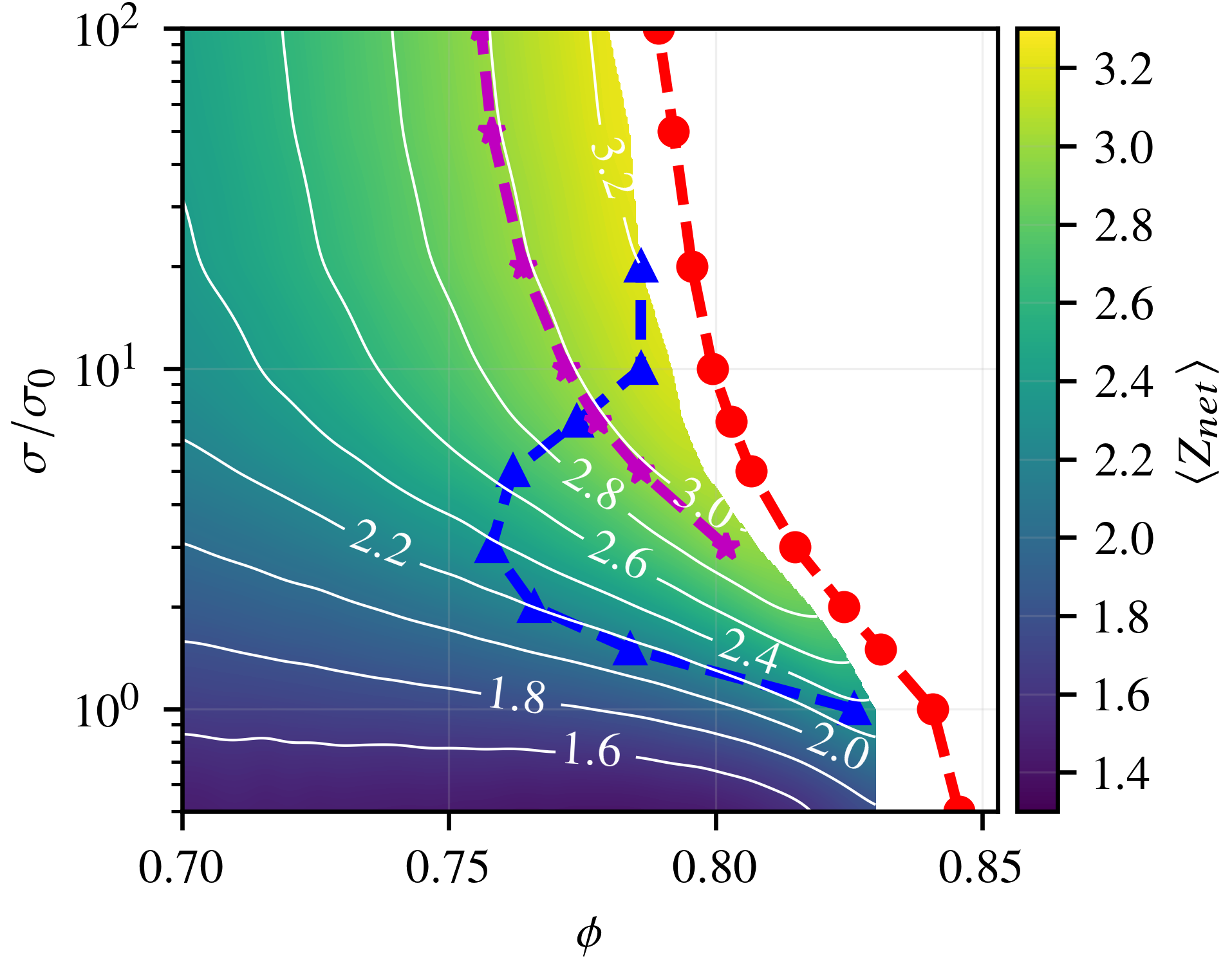}
    \caption{Contour plot of the mean contact number $\langle Z_{net}\rangle$. White curves are iso-$\langle Z_{net}\rangle$ contours, and the blue, red, and magenta lines match those in the flow diagram. For the present high-friction system, the line $\phi_c(\sigma)$ passes close to the $\langle Z_{net}\rangle = 3$ isoline.}
    \label{fig:znet_contour}
\end{figure}

As shown in \cref{fig:znet_contour}, the mean frictional contact number $\langle Z_{net}\rangle$, computed as the average contact number over all particles in the contact network, increases with both stress and solid fraction. This is the expected consequence of increasing stress progressively overcoming the short-range repulsion and converting lubricated near-contacts into frictional constraints \citep{seto2013,mari2014,mari2015,wyart2014,morris2018,guy2018,singh2022,thomas2018}. Each contact constrains relative motion, so an increasing frictional contact number is equivalent to more restrictions on local degrees of freedom \citep{guy2018,singh2022,vandernaald2024}. However, contact number alone does not determine whether those constraints are arranged to produce rigidity or long-range correlations.

The iso-$\langle Z_{\mathrm{net}}\rangle$ contours are not horizontal in the $(\sigma/\sigma_0,\phi)$ plane: at larger stress they bend and become nearly vertical, indicating that once frictional contacts are activated, solid fraction strongly influences how many such contacts the suspension can sustain. This observation should not be interpreted as a departure from the Wyart--Cates (WC) model \citep{wyart2014}, whose stress-dependent state variable $f$ is the fraction of contacts that are frictional, rather than the mean frictional coordination number $\langle Z_{\mathrm{net}}\rangle$. Even when $f$ is governed primarily by $\sigma/\sigma_0$, $\langle Z_{\mathrm{net}}\rangle$ can retain a dependence on $\phi$ through the number and organization of contacts available at a given packing. The high-stress, large-$\phi$ corner of the state diagram is also where mesoscale contact-network topology becomes relevant in addition to the overall fraction of constrained contacts \citep{edens2021,sedes2022,lemaitre2021,damico2025,sharma2026}.

For the present $\mu=100$ system, the critical line $\phi_c(\sigma)$ passes close to $\langle Z_{\mathrm{net}}\rangle=3$. Because sliding contacts are infrequent at this large friction coefficient, each frictional contact generally provides both a normal and a tangential constraint; constraint counting therefore gives the large-friction isostatic coordination $Z_{\mathrm{iso}}=d+1=3$ in two dimensions \citep{unger2005,shundyak2007}. The proximity of the independently determined rigidity-percolation transition to this coordination is consistent with the onset of a finite jamming probability at $\phi_c$ found previously for sufficiently stiff particles \citep{pandare2026}. It is, however, an empirical feature of the present $\mu=100$ data rather than a universal coordination criterion. We have not tested whether the same alignment persists for $\mu<1$, where fully mobilized contacts enter a generalized isostaticity condition \citep{shundyak2007}. In three dimensions the large-friction isostatic value shifts to $Z_{\mathrm{iso}}=4$; the contact-number map retains the same qualitative structure, with values exceeding~4 in the high-stress, large-$\phi$ region (Supplemental Material, Fig.~S2).

\begin{figure}[!t]
    \centering
    \includegraphics[width=\singlefigwidth]{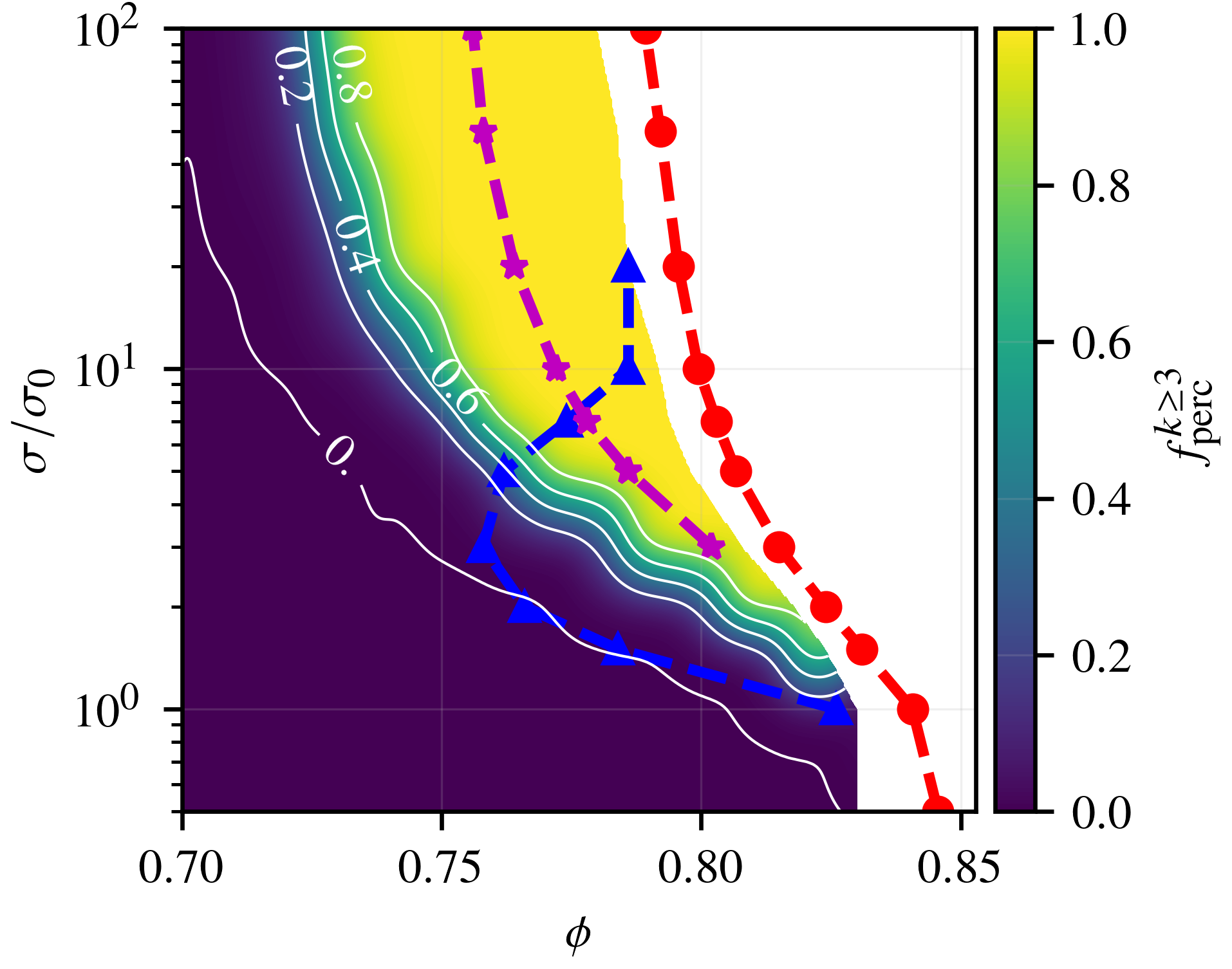}
    \caption{Contour plot of the fraction of samples having system-spanning $k\ge 3$ neighbor cluster. White curves are iso-$f_{\mathrm{perc}}^{k\ge 3}$ contours. Percolation first appears on the lower branch of the DST onset line and then continues to lower $\phi$ (below the DST region) and higher stress; thus the $f_{\mathrm{perc}}^{k\ge 3}=0$ contour becomes nearly vertical and roughly parallel to the line $\phi_c(\sigma)$  in the high-stress regime.}
    \label{fig:contours_f_perc_k3}
\end{figure}

The $k$-neighbor analysis of \cref{fig:contours_f_perc_k3} provides another connection between network structure and $\langle Z_{net}\rangle$. Following \citet{goyal2024}, particles are first identified by having at least $k$ frictional contacts and are then grouped into clusters by contact connectivity. \citet{goyal2024} analyzed $k\ge 4$ clusters because that threshold isolates locally overconstrained particles in the 3D case they studied; by dimensional analogy, the corresponding threshold for the present 2D monolayers is $k\ge 3$, which identifies locally isostatic or overconstrained particles. This interpretation is consistent with recent 2D minimal-model simulations showing that thickening states contain distinct contact-network building blocks and large spanning assemblies \citep{buchholtz2026}. In the present data, constrained clusters first span the system on the lower branch of the DST onset and then the line for their onset of percolation continues to lower $\phi$ and higher stress, largely following the contours of the $\phi_c(\sigma)$ and shear-jamming line. Thus, the contact number captures the growth of the constraint network, while the $k\ge 3$ percolation analysis indicates where that network becomes system-spanning in the suspension shear flow studied. The analogous $k\ge 4$ and $k\ge 5$ percolation fractions in three dimensions show a similar state-diagram pattern (Supplemental Material, Figs.~S3(a) and~S3(b)). Note that the onset of $k\ge 3$ percolation occurs at significantly lower $\phi$ than the onset of rigidity at $\phi_c(\sigma)$. The next sections consider the mechanical significance of these contact structures by studying their relation to velocity fluctuations and their correlations.

\section{Spatial correlations of translational velocities}\label{sec:vcorr}

We consider spatial correlations of $\tilde{v}_x = v_x - \dot\gamma y$, where $\dot{\gamma} y$ is the streamwise affine velocity. Velocity fluctuations, particle dispersion, and shear-induced diffusion carry information about microstructural organization in sheared suspensions \citep{brady1997,drazer2004,pham2015,souzy2015,denn2014,athani2024}, and particulate systems approaching arrest develop dynamical heterogeneity and cooperative motion \citep{biswas2026}. The question here is whether frictional contacts near DST give rise to correlated particle motion and over what length scale. The gradient-direction component $C_{\tilde{v}_y}(r)$ shows the same qualitative dependence on stress and solid fraction but with larger statistical noise, and so we focus on $C_{\tilde{v}_x}(r)$.

\begin{figure}[!t]
    \centering
    \includegraphics[width=\singlefigwidth]{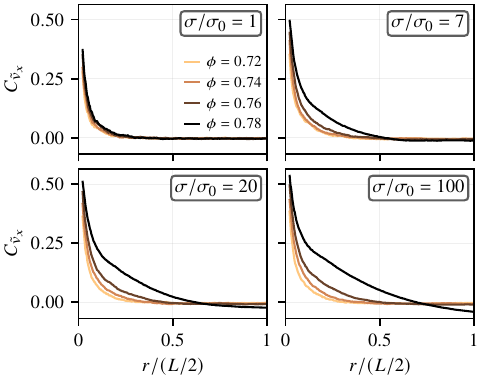}
    \caption{Spatial correlation function of the streamwise non-affine velocity $\tilde{v}_x$ for representative systems.}
    \label{fig:spatial_vx_corr_representative}
\end{figure}

We compute the equal-time, variance-normalized spatial correlation
\begin{equation}
    C_{\tilde{v}_x}(r) = \frac{\langle \tilde{v}_{x,i}\,\tilde{v}_{x,j} \rangle_r}{\langle \tilde{v}_x^2 \rangle}\ ,
    \label{eq:Cvx}
\end{equation}
where $\langle\cdot\rangle_r$ averages over particle pairs at separation $r = |\mathbf{x}_j-\mathbf{x}_i|$, with $\mathbf{x}_{i}$ the center position of particle $i$; we characterize the curve by its zero-crossing distance $r_0$. Representative $C_{\tilde{v}_x}(r)$ curves are shown in \cref{fig:spatial_vx_corr_representative}; $r_0$ increases with both $\sigma/\sigma_0$ and $\phi$.  The largest values occur in the region in $(\sigma/\sigma_0, \phi)$ parameter space where \cref{fig:flow_diagram,fig:znet_contour,fig:contours_f_perc_k3} show rapid growth of frictional contacts, proximity to the critical transition, and the appearance of system-spanning constrained clusters (Supplemental Material, Fig.~S9).

\begin{figure}[!t]
    \centering
    \includegraphics[width=\singlefigwidth]{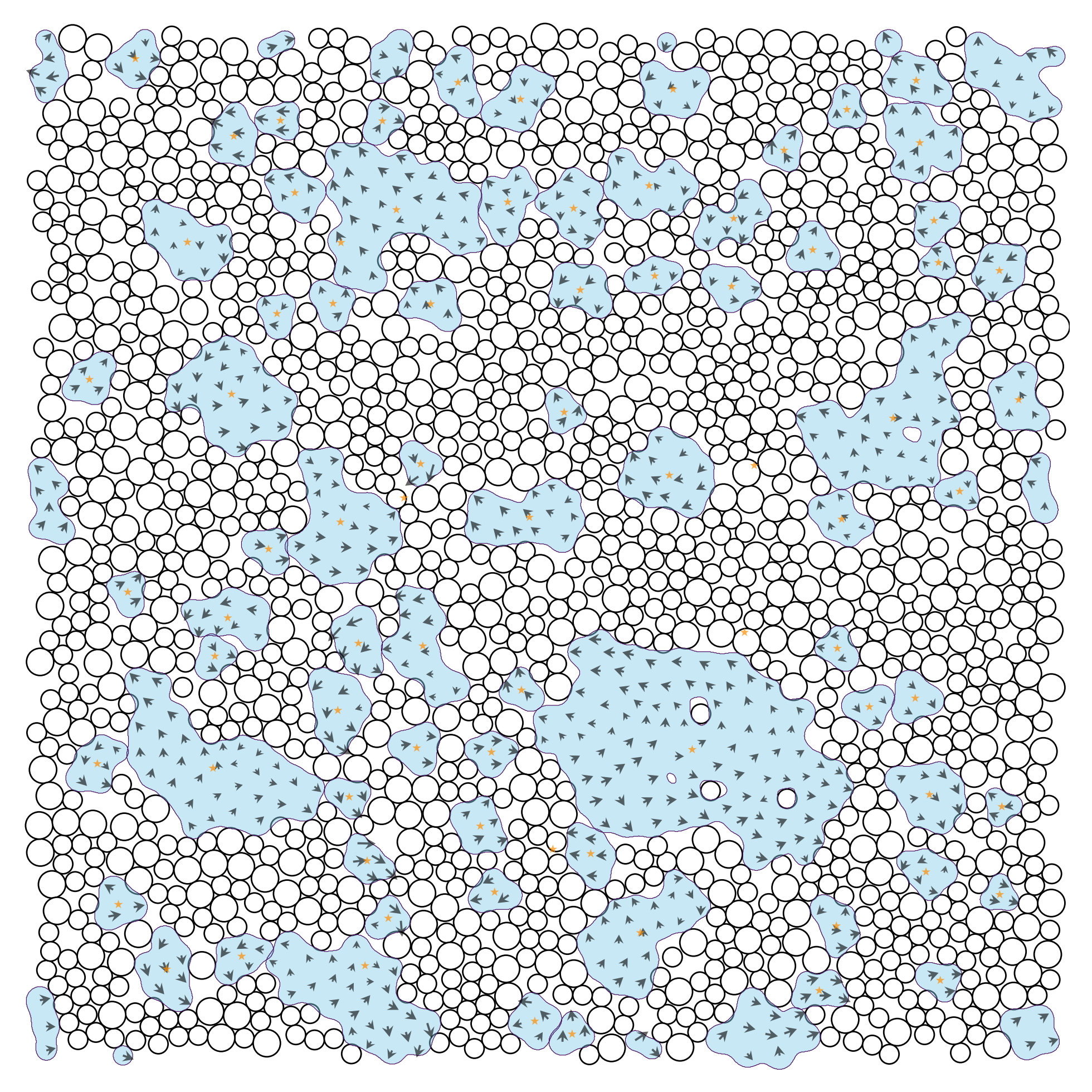}
    \caption{Representative snapshot at $\sigma/\sigma_0=100$ and $\phi=0.756$. Blue contours identify pebble-game rigid clusters, arrows show the translational velocities of particles in rigid clusters, and orange $\star$ symbols mark the cluster centers of mass.}
    \label{fig:coarse_grained}
\end{figure}

\begin{figure*}[!t]
    \centering
    \includegraphics[width=\textwidth]{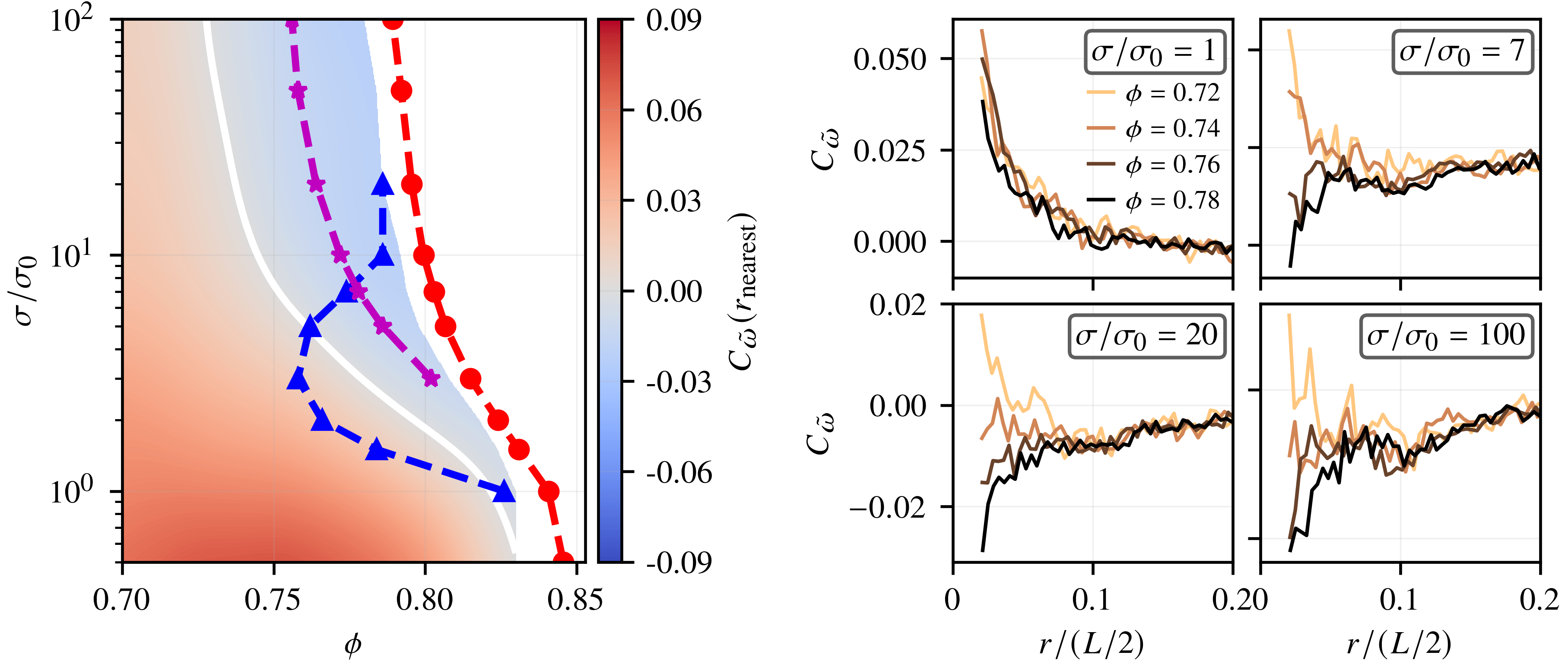}
    \caption{Rotational correlations. Left: near-contact rotational correlation in the $(\sigma/\sigma_0\,,\,\phi)$ plane, with the solid white line indicating the zero-isoline. Positive values indicate correlated non-affine angular velocities, and negative values indicate anti-correlated non-affine angular velocities. Right: spatial correlation function of the non-affine rotational velocity $\tilde{\omega}_{y}$ at representative state points; each panel shows results at a fixed stress, and line colors indicate $\phi$ as shown in the legend.}
    \label{fig:contours_omy_nearest}
\end{figure*}

This comparison does not establish a scaling relation between the correlation length and rigid-cluster size or distance from the critical line. It is, however, consistent with a simple physical picture: once frictional contacts become numerous enough to impose an extended network of kinematic constraints, nearby particles are less free to fluctuate independently, and coherent translational motion persists over larger distances. This is consistent with simulation work linking shear-induced diffusivity and viscosity to the size of collective motions or rigid-cluster-like domains in frictional suspensions \citep{singh2023,santra2025,pandare2026}.

The representative configuration in \cref{fig:coarse_grained} shows that the blue-outlined rigid clusters translate coherently over several particle diameters while also displaying collective rotation. The conditional angular-velocity distributions in \cref{fig:angVel_rig_nonrig} add a state-dependent contrast: at lower solid fraction, particles outside small rigid clusters accommodate more relative rotational motion, whereas the two population distributions are comparable at $\phi_c$. As solid fraction and stress increase further, rigid domains grow and ultimately form system-spanning clusters that incorporate the majority of particles (Supplemental Material, Fig.~S8). The remaining particles outside rigid clusters then become geometrically confined, and their motion relative to the rigid population is suppressed, as reflected by their narrow angular-velocity distribution at high solid fraction in \cref{fig:angVel_rig_nonrig}. This evolution is consistent with the growing translational correlation length, but the angular-velocity distributions do not by themselves quantify how the local translational velocity gradient is partitioned between the two populations.

\section{Rotational velocity}\label{sec:omy}

Rotational motion complements the translational picture. Whereas $\tilde{v}_x$ and its correlation $C_{\tilde{v}_x}(r)$ capture the spatial extent of collective motion, the non-affine angular velocity $\tilde{\omega} = \omega_y - 0.5\dot{\gamma}(t)$, normalized by $\dot{\gamma}(t)$ to give $\omega' = \tilde{\omega}/\dot{\gamma}$, probes how neighboring particles rotate relative to the affine background. In dense suspensions, tangential relative motion at or near contact couples translation and rotation, so rotational fluctuations are naturally linked to the same frictional constraints and relative motions that govern thickening \citep{seto2013,mari2014,morris2020,singh2022}.

Here we focus on $\tilde{\omega}_{y}$, the fluctuation about the affine rotation, as a local probe: tangential motion at contacts couples particle translation and rotation through the same frictional constraints that drive thickening. \cref{fig:contours_omy_nearest} shows that the near-contact rotational correlation $C_{\tilde{\omega}_{y}}(r_{\mathrm{nearest}})$ becomes increasingly negative in the dense, high-stress region. Because this correlation is computed from $\tilde{\omega}_y$, positive and negative values denote correlated and anti-correlated non-affine rotational fluctuations, respectively; a negative value does not necessarily imply opposite laboratory-frame angular velocities. For an isolated pair undergoing pure rolling in the absence of a background flow, vanishing relative tangential motion of the particle centers requires opposite particle rotations. In uniform simple shear, however, symmetry of the isolated-pair interaction precludes this anti-rotation, and the particles instead share a common angular velocity as the pair undergoes solid-body rotation about the contact. The sign change in \cref{fig:contours_omy_nearest} is a magnitude-weighted average over a mixture of multi-contact environments. It therefore indicates the increasing dominance of anti-correlated non-affine rotational fluctuations at short range, but does not uniquely identify the topology or rigidity of the contributing contact network.

\subsection{Statistics}\label{sec:omypdf}

\begin{figure*}[!t]
    \centering
    \includegraphics[width=\textwidth]{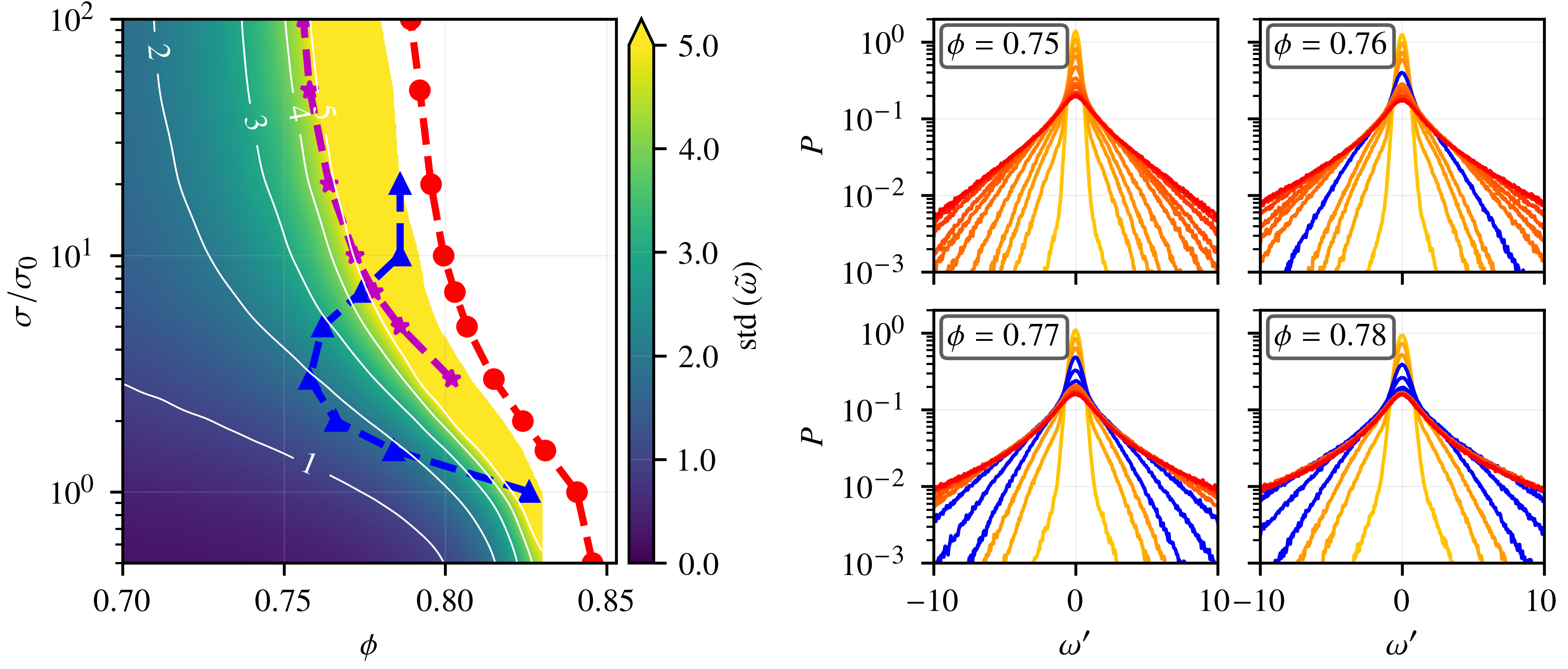}
    \caption{Statistics of the non-affine rotational velocity. Left: standard deviation of $\tilde{\omega}$ across the $(\sigma/\sigma_0\,,\,\phi)$ plane. Right: probability density functions of $\omega' = \tilde{\omega}/\dot{\gamma}$ at four solid fractions; each panel shows all simulated stress values. Warm curves (yellow to dark red, in order of increasing stress) indicate states classified rheologically as lying below the DST onset at that $\phi$; blue curves indicate states in the DST region.}
    \label{fig:Pomy}
\end{figure*}

The right panels of \cref{fig:Pomy} show $P(\omega')$ at four solid fractions, with each panel spanning the full simulated stress range. Across the full sweep, increasing stress and solid fraction generally broadens the distribution, lowering its central peak and enhancing its tails. The blue curves identify state points in the DST region, but the PDFs evolve gradually and display no abrupt or unique change at the DST boundary. Because $\tilde{\omega}_{y}$ is defined relative to the affine rotation, this broadening reflects growing non-affine rotational fluctuations rather than the imposed background shear itself. The distributions also depend on the contact law: the present simulations include sliding friction but no independent rolling resistance, and the resulting $P(\omega')$ should therefore not be interpreted as universal for systems with different rotational constraints.

The left panel of \cref{fig:Pomy} summarizes this trend across the full $(\sigma/\sigma_0\,,\,\phi)$ plane. The standard deviation of $\tilde{\omega}_{y}$ grows toward large stress and large $\phi$, roughly following the contact-number contours of \cref{fig:znet_contour}, with the strongest values developing in the critical and shear-jamming region. Stress and solid fraction therefore both matter: increasing stress activates frictional contacts and constraints, while increasing $\phi$ amplifies the rotational fluctuations once the system is sufficiently constrained. A comparison with \cref{fig:spatial_vx_corr_representative} shows that the rotational and translational observables strengthen in broadly the same region, but they measure different aspects of the motion. The correlation length $r_0$ characterizes the spatial extent of collective translational motion, whereas $\mathrm{std}\left(\tilde{\omega}_{y}\right)$ measures the intensity of local rotational fluctuations. \cref{fig:contours_omy_nearest} indicates whether neighboring particles have correlated or anti-correlated non-affine rotational fluctuations. The findings here are consistent with numerical work linking friction-driven diffusivity growth to a broadening of rotational-velocity variance \citep{zhang2024}, and with recent rigidity-based analyses of the same region \citep{santra2025,pandare2026}.

\begin{figure}[!t]
    \centering
    \includegraphics[width=\singlefigwidth]{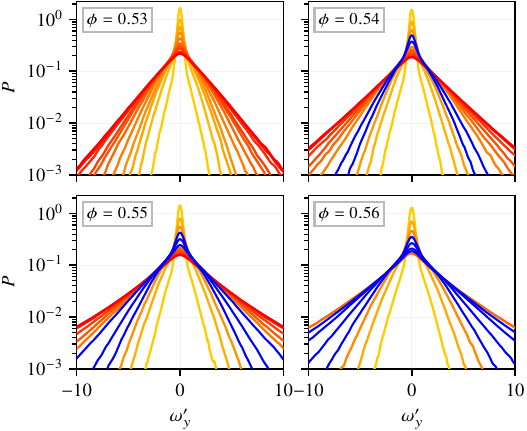}
    \caption{Three-dimensional rotational statistics. Probability density functions of the normalized non-affine angular velocity $\omega'_y$ in 3D simulations at representative state points broaden with increasing stress and solid fraction, consistent with the trend observed in the 2D monolayer; note that here $\phi$ is particle volume fraction.}
    \label{fig:3d_pomy}
\end{figure}

\cref{fig:3d_pomy} shows that three-dimensional suspensions have similar rotational statistics: the $\omega'_y$ distributions in 3D simulations exhibit the same gradual broadening with increasing stress and solid fraction as in the 2D monolayer. The 3D data are used only for this qualitative comparison; 2D is the primary focus here  because it also allows the pebble-game rigidity analysis. Standard-deviation and excess-kurtosis maps of $\tilde\omega_y$ in three dimensions confirm the same state-diagram trend (Supplemental Material, Figs.~S4(a) and~S4(b)), as do the PDFs swept over stress at fixed $\phi$ as shown in \cref{fig:3d_pomy}. All three Cartesian components of the non-affine rotation exhibit fluctuations that broaden equivalently at fixed stress and solid fraction (Supplemental Material, Fig.~S5).

\subsection{Particle rotation inside and outside rigid clusters}\label{sec:omyrig}

\begin{figure}[!t]
    \centering
    \includegraphics[width=\singlefigwidth]{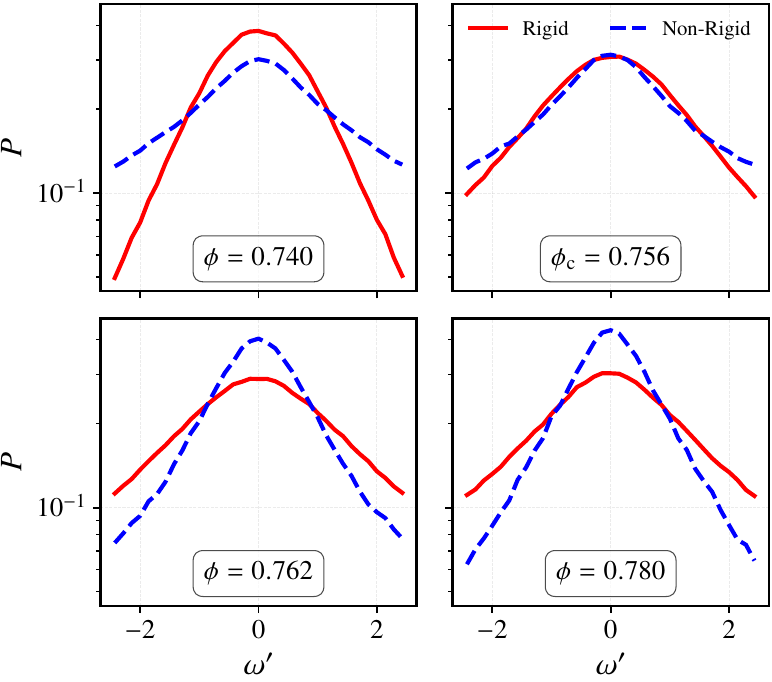}
    \caption{Conditional angular-velocity distributions for particles inside and outside rigid clusters at $\sigma/\sigma_0=100$. The two population PDFs are normalized separately. The plotted variable is the normalized non-affine angular velocity $\omega'=\tilde{\omega}(t)/\dot{\gamma}(t)$, where $\tilde{\omega}=\omega(t)-0.5\dot{\gamma}(t)$. The uncertainty envelopes quantify sampling variability over the steady-state data. As solid fraction increases, the distribution inside rigid clusters broadens, while the distribution outside them narrows above the critical solid fraction $\phi_c=0.756$.}
    \label{fig:angVel_rig_nonrig}
\end{figure}

Membership in a rigid cluster is identified from the pebble-game construction of Sec.~\ref{sec:pebblegame}; all simulated particles otherwise share the same material model. Within a rigid cluster, contacts constrain relative motion and favor coherent translation over several particle diameters; this collective motion is consistent with the growing translational correlations of Sec.~\ref{sec:vcorr} and with experimental observations of correlated motion near jamming \citep{biswas2026}.

\begin{figure}[!t]
    \centering
    \includegraphics[width=\singlefigwidth]{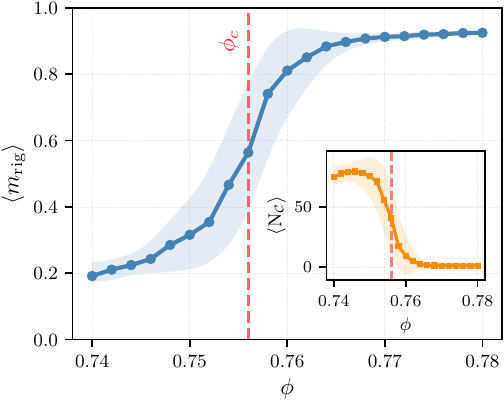}
    \caption{Rigid-cluster statistics at $\sigma/\sigma_0=100$. The main panel shows the mean rigid fraction $\langle m_{\mathrm{rig}}\rangle$, and the inset shows the mean number of rigid clusters $\langle N_{\mathcal C}\rangle$. Each point is averaged over steady-state configurations, and the shaded regions indicate one standard deviation. The vertical dashed line marks the critical solid fraction $\phi_c=0.756$ for this stress.}
    \label{fig:cluster_stats}
\end{figure}

Here, we compare the conditional distributions $P(\omega'\mid\mathrm{inside})$ and $P(\omega'\mid\mathrm{outside})$ for particles inside and outside rigid clusters at fixed stress; each is normalized independently within its own population. At $\phi=0.740$, $\langle m_{\mathrm{rig}}\rangle\simeq0.19$, corresponding to approximately 380 rigid-cluster particles and 75 rigid clusters per configuration, or roughly five particles per cluster on average (\cref{fig:cluster_stats}); the nearby strain-resolved data show that a comparable rigid fraction persists during steady-state sampling (Supplemental Material, Fig.~S7). Thus, the rigid-cluster population in \cref{fig:angVel_rig_nonrig} at $\phi=0.74$ is appreciable, although it consists primarily of small, isolated clusters. The uncertainty envelopes show that the separation between the two conditional PDFs exceeds the sampling variability. At lower $\phi$, the distribution inside rigid clusters is more concentrated near $\omega'=0$ than the distribution outside them. This is consistent with the internal constraints of small rigid clusters favoring collective motion, while particles outside rigid clusters retain unconstrained modes and accommodate more relative motion; the figure does not identify a unique mechanism. At $\phi_c=0.756$, the two distributions are comparable, while above this critical solid fraction, the distribution outside rigid clusters narrows again, consistent with increasing geometric confinement by the surrounding rigid domains. The distribution inside rigid clusters remains broader, reflecting particle-scale rotation and cluster solid-body motion associated with increasingly large constrained structures. In 3D simulations, where the pebble-game construction is not available, we classify particles by contact count following the $k$-neighbor approach of \citet{goyal2024}. For both $k\ge 4$ and $k\ge 5$ thresholds, the constrained population has a narrower rotational-velocity distribution than the remainder at high stress (Supplemental Material, Fig.~S6), consistent with the contrast between particles inside and outside rigid clusters observed in 2D at lower packing fractions.

\begin{figure}[!t]
    \centering
    \includegraphics[width=\singlefigwidth]{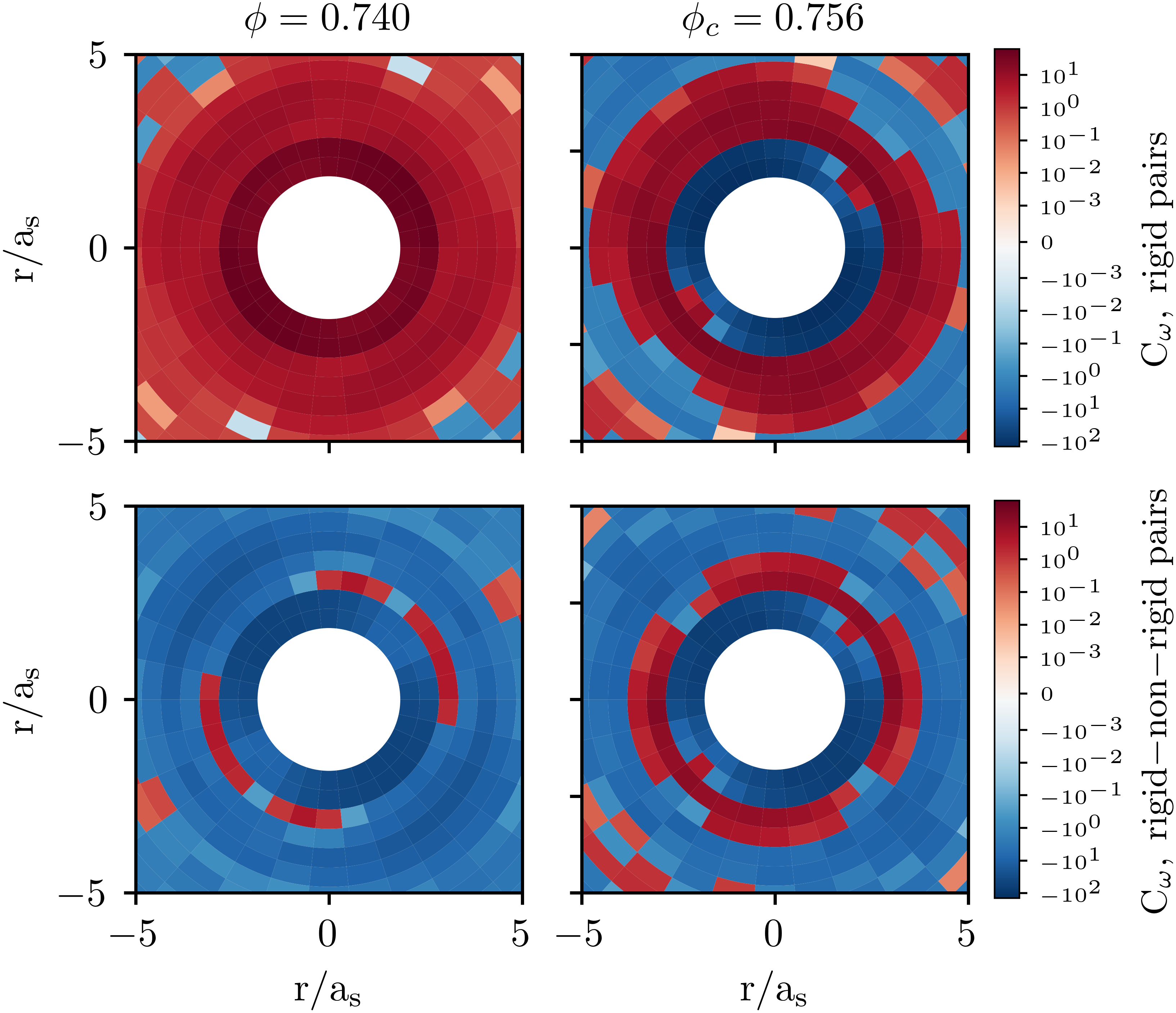}
    \caption{Correlation of the non-affine angular velocity, $C_{\omega}(r,\theta)$, for pairs of particles inside rigid clusters (top row) and pairs with one particle inside and one outside a rigid cluster (bottom row) at $\sigma/\sigma_0=100$. The left column shows $\phi=0.740$, and the right column shows the critical solid fraction $\phi_c=0.756$. Red sectors indicate correlated non-affine angular velocities, and blue sectors indicate anti-correlated non-affine angular velocities. All four panels use the same symmetric-logarithmic normalization and a single shared colorbar. The color limits are symmetric about zero and set by the largest absolute value of $C_\omega$ across the four panels. The axes represent the separation $r$ normalized by the small-particle radius $a_s$. The shear flow is from left to right at the top of each image and from right to left at the bottom.}
    \label{fig:angvel_correlations}
\end{figure}

To resolve where these contrasts in rotational motion are concentrated in microstructural terms, we compute angular-velocity correlations as a function of the vector separation, i.e. considering both $r$ and the angle $\theta$ relative to the flow direction. For particle pairs drawn from a chosen population combination, the correlation is
\begin{equation}
    \begin{split}
        C_{\omega}(r,\theta) &=
        \left\langle \omega^\prime_{i}\omega^\prime_{j}\right\rangle\ ,
    \end{split}
\end{equation}
where $r$ and $\theta$ are the distance and angle between the centers of particles $i$ and $j$.

\cref{fig:angvel_correlations} separates pairs in which both particles lie inside rigid clusters from pairs with one particle inside and one outside a rigid cluster. At lower solid fraction, pairs inside rigid clusters mostly have positively correlated non-affine angular velocities over the first few particle diameters, as expected for small coherent clusters. Near the critical solid fraction, the first coordination layer becomes more heterogeneous and a stronger anti-correlated region appears. Pairs spanning the inside--outside populations are predominantly anti-correlated at short range, while a secondary positive shell appears at higher $\phi$, indicating spatially structured non-affine rotational correlations. These population-resolved correlations show where rotational contrasts are concentrated without assigning a unique network topology to the unconditioned correlation in \cref{fig:contours_omy_nearest}.

\cref{fig:cluster_stats} quantifies how the rigid-cluster population evolves with solid fraction at $\sigma/\sigma_0=100$. The mean rigid fraction $\langle m_{\mathrm{rig}}\rangle$ increases with solid fraction and approaches unity at the largest $\phi$, while the number of distinct clusters $N_{\mathcal{C}}$ decreases. The mean cluster size therefore grows rapidly, as  the rigid population evolves from many small isolated clusters to a small number of large domains. For example, the configuration in \cref{fig:coarse_grained} has $m_{\mathrm{rig}}=0.33$ and 80 rigid clusters; the largest cluster contains 92 particles, while the mean cluster size is around 8. To aid in interpretation of the image, recall that it is periodically replicated in the simulation technique.

\begin{figure}[!t]
    \centering
    \includegraphics[width=\singlefigwidth]{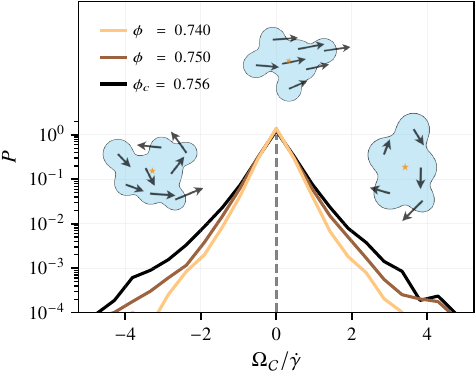}
    \caption{Coarse-grained cluster angular-velocity distributions at $\sigma/\sigma_0=100$ for varying $\phi$. The quantity $\Omega_{\mathcal{C}}$ (\cref{eq:coarse-grained}) captures the rotational velocity of each cluster about its center of mass, distinct from the individual particle rotations of \cref{fig:angVel_rig_nonrig}. Three representative clusters illustrating rotational motion are shown. The arrows represent the non-affine translational velocity of the rigid particles. The cluster with net counter-clockwise rotation is on the left ($\Omega_{\mathcal{C}} < 0$), and that with clockwise rotation is on the right ($\Omega_{\mathcal{C}} > 0$), with the center of mass marked by an orange $\star$ symbol. A rigid cluster has zero net rotational motion when particles are stationary or have only translational velocities; the latter case is illustrated in the center cluster. Data above $\phi=0.756$ are not shown because too few distinct clusters remain for reliable statistics (Supplemental Material, Figs.~S8(c) and~S8(f)).}
    \label{fig:coarse-grained_angvel}
\end{figure}

Beyond individual particle rotation, correlated translational velocity fluctuations within a rigid cluster can produce rotation about the cluster center of mass. For a rigid cluster $\mathcal{C}$, we define the coarse-grained angular velocity
\begin{equation}
    \Omega_{\mathcal{C}} =
    \frac{\sum_{i\in\mathcal{C}}\left[(\mathbf{r}_i-\mathbf{R}_{\mathcal{C}})\times(\mathbf{v}_i-\mathbf{V}_{\mathcal{C}})\right]_y}
    {\sum_{i\in\mathcal{C}}\left|\mathbf{r}_i-\mathbf{R}_{\mathcal{C}}\right|^2}\ ,
    \label{eq:coarse-grained}
\end{equation}
where $\mathbf{R}_{\mathcal{C}}$ and $\mathbf{V}_{\mathcal{C}}$ are the cluster center-of-mass position and velocity. This quantity measures cluster-scale angular motion and is distinct from the individual particle rotation rate $\tilde{\omega}_y$. Its distributions at $\sigma/\sigma_0=100$ are shown in \cref{fig:coarse-grained_angvel}. These distributions differ markedly from the particle-scale angular velocities of \cref{fig:angVel_rig_nonrig}: they broaden with increasing $\phi$ until the critical fraction, then narrow as system-spanning rigid clusters form, a trend distinct from the individual particle rotation behavior. Three representative clusters are shown along with the translational velocity vectors (black arrows) and the center of mass (orange star). The cluster on the left has a net anti-clockwise motion around the center of mass, with $(\Omega_{\mathcal{C}} < 0)$, whereas the cluster on the right shows a net clock-wise motion $(\Omega_{\mathcal{C}} > 0)$. Solid fractions above the critical line are dominated by large rigid domains that increasingly resist the imposed flow, driving the cluster angular velocity toward $\Omega_{\mathcal{C}}=0$. However, at $\phi < \phi_c$ where smaller clusters dominate, we often observe rigid particles with aligned translational velocities that do not display significant rotational motion, a case illustrated by the representative rigid cluster in the center in \cref{fig:coarse-grained_angvel}.

\section{Conclusions}\label{sec:conclusions}

We have compared structural and kinematic observables across the flow-state diagram for dense shear-thickening suspensions, showing how behavior depends on stress and solid fraction. We have found that the growth of frictional contact-network organization is accompanied by systematic changes in the fluctuational motions undergone by highly-concentrated suspended particles in a bulk simple-shear flow.

Three structural observables were examined. In order of increasing organizational complexity, the first is the mean frictional contact number $Z_{\mathrm{net}}$, which grows with both stress and solid fraction; $Z_{\mathrm{net}}$ contours bend from roughly horizontal at moderate stress to nearly vertical at large stress, indicating that solid fraction continues to control the number of sustainable frictional constraints even after the lubricated-to-frictional transition is well under way. Second, the $k\ge 3$ percolation analysis shows that system-spanning isostatic or overconstrained contact structures first appear along the lower branch of the DST onset and the percolation of these structures continues to lower $\phi$ at elevated stress, eventually paralleling $\phi_c$ as $\sigma/\sigma_0$ becomes large. Third, the pebble-game rigidity analysis provides a more detailed view: rigid clusters emerge below shear jamming, and their size distribution shifts from many small isolated domains to highly fluctuating in size, occasionally dropping to just a few large potentially spanning objects as the critical solid fraction is approached at fixed stress. These three descriptions are complementary: the contact number quantifies the density of constraints, the $k$-neighbor percolation identifies where those constraints form a system-spanning backbone, and the rigidity analysis identifies the subsets of that backbone that are minimally rigid.

The kinematic observables evolve systematically with increasing stress and solid fraction in the region that contains DST and approaches shear jamming. The translational velocity correlation length grows with stress and solid fraction, and its largest values occur where $Z_{\mathrm{net}}$, $k\ge 3$ percolation, and rigid-cluster fluctuations are also large, consistent with an extended constraint network sustaining coherent motion over longer distances. The rotational statistics provide a more local view: the standard deviation of $\tilde{\omega}_y$ broadens gradually, while near-contact correlations change from weakly positive to increasingly negative, indicating stronger anti-correlation between neighboring particles' non-affine rotational fluctuations. These trends overlap the DST region but do not exhibit an abrupt or unique signature at its boundary. Conditioning on rigid-cluster membership adds further detail: above $\phi_c$, the distribution outside rigid clusters narrows while that inside rigid clusters continues to broaden.

Connectivity, rigidity, and velocity correlations are therefore complementary but distinct indicators of the progressive organization of the contact network and particle motion that accompanies shear thickening and the approach to jamming.

The most direct extension is to quantify how the velocity-correlation length scales with rigid-cluster size near the criticality line, and to test whether that scaling is consistent with the continuous transition identified from rigid-cluster size fluctuations \citep{santra2025,pandare2026}. The contrast between particles inside and outside rigid clusters in both translational and rotational statistics also raises the question of how each population contributes to the overall viscosity; a dissipation analysis separating their contributions to the hydrodynamic stress \citep{essayah2022} would connect this kinematic picture to the macroscopic rheology. More broadly, the co-evolution of contact-network structure and particle motion shown here may help interpret the localized and transient jammed regions observed experimentally in thickening suspensions \citep{rathee2020,barik2024,moghimi2024,moghimi2025}, although simulations are limited to exploring behavior for a much smaller number of particles than these experiments.

\section*{Acknowledgments}

JFM, BAP, and RP were supported by NSF DMR-2410984 and CBET-2228680. BC was supported by NSF CBET-2228681. Discussions with Prof. E. Del Gado (Georgetown University) were very helpful.

\section*{Author declarations}

The authors declare no competing interests.

\section*{Data availability}

The data that support the findings of this study are available upon reasonable request.

\interlinepenalty=10000
\clubpenalty=10000
\widowpenalty=10000
\displaywidowpenalty=10000
\bibliography{zotero}

\end{document}